\documentclass[a4paper,12pt]{book}

\usepackage{lscape}

\usepackage{tikz}

\usepackage{geometry}

\usepackage{graphicx}

\usepackage{verbatim}

\usepackage{hyperref}

\usepackage{euscript}

\usepackage{amssymb,amsmath}

\usepackage{appendix}

\makeatletter
\renewcommand{\@makechapterhead}[1]{%
\vspace*{50 pt}%
{\setlength{\parindent}{0pt} \raggedright \normalfont
\bfseries\Huge
\ifnum \value{secnumdepth}>1
   \if@mainmatter\thechapter.\ \fi%
\fi
#1\par\nobreak\vspace{40 pt}}}
\makeatother

\usepackage{feynmf}

\begin{document}


\begin{titlepage}

\begin{center}

\includegraphics[width=100mm, height=25mm]{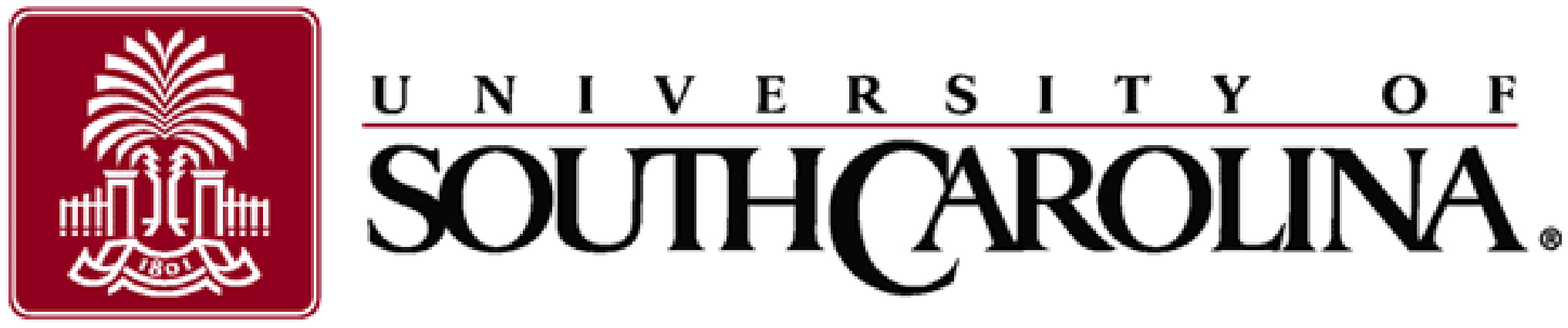}\\
\ \\ 
\textsc{\Large
South Carolina Honors College\\
\ \\
Senior Thesis\\
\ \\}
%
%
{\huge \bfseries
The Weak Force: \\ From Fermi to Feynman
\\}
\ \\
\ \\
\begin{minipage}{0.4\textwidth}
\begin{flushleft} \large
\emph{Author:}\\
Alexander Lesov\\
\end{flushleft}
\end{minipage}
%
\begin{minipage}{0.4\textwidth}
\begin{flushright} \large
\emph{Mentor:} \\
Dr. Milind Purohit\\
\footnotesize{\ }
\end{flushright}
\end{minipage}
\vfill 
%
{\large \today}
\end{center}

\end{titlepage}

\tableofcontents

\chapter*{Thesis Summary}

\addcontentsline{toc}{chapter}{Thesis Summary}

\begin{quotation}
 \textit{“I think physicists are the Peter Pans of the human race. They never grow up and they keep their curiosity.”} - I.I. Rabi
\end{quotation}

\paragraph*{}\begin{Large}$\mathcal{I}$\end{Large}f one was to spend a few minutes observing the physical phenomena taking place all around them, one would come the conclusion that there are only two fundamental forces of nature, Gravitation and Electromagnetism. Just one century ago, this was the opinion held by the worldwide community of physicists. After many decades spent digging deeper and deeper into the heart of matter, two brand new types of interactions were identified, impelling us to add the Strong and Weak nuclear forces to this list. Of these two new forces, the Weak has done the most to shatter long-held beliefs and to ultimately guide us towards a deeper understanding of our universe. 

In this work we will trace the history of our understanding of the weak force from early observations and explanations of $\beta$ decay all the way to the doorstep of Electroweak unification, one of the greatest accomplishments of modern physics. In the process we will encounter some of the greatest minds of the last century and try to understand the significance of some of their work.

In the first chapter we examine the first attempts to construct a theory of the original weak interaction, $\beta$ decay. The first section is dedicated to an explanation of the problem of a continuous electron energy spectrum in such interactions. The second gives a brief biographical sketch of one of the most interesting characters of modern physics, Wolfgang Pauli. In the third section we discuss Pauli's proposal of a new particle in order to salvage energy conservation and alleviate the problem of ``wrong statistics''. Sections four and five are dedicated to mathematical descriptions of early quantum theory and Enrico Fermi's original theory of $\beta$ decay, constructed in analogy to the newly created theory of Quantum Electrodynamics, respectively.

The second chapter discusses the discovery of one of the most striking features of the Weak force, its violation of reflection symmetry. We spend the first section examining the general idea of a symmetry and its connection, through Noether's theorem, to the concept of a conservation law. Specifically, we examine the relationship between parity conservation and reflection symmetry. Finally, we consider an early observation of parity conservation in the form of Laporte's rule. In the second section we briefly examine the lives of T.D. Lee and C.N. Yang, two of the most important physicists in our story. In the third section we examine a puzzle which led some, including Lee and Yang, to begin questioning parity conservation. The next section gives an account of the 1956 Rochester High Energy Conference, at which the possibility of a parity violating weak force was first brought up in a serious context. In the last sections we examine a wonderful experiment, conducted by Madame Wu, which confirmed this suspicion.

The final chapter attempts to generalize the Fermi theory which was originally developed in the first chapter into a more accurate form through the use of results discussed in the second chapter. The first section generalizes Fermi's original theory through the explanation of the concept of covariant forms. The second discusses the notion of a universal Fermi interaction and includes a very short biography of its originator, Bruno Pontecorvo. The next section, which is split into two subsections, examines the consequences of the parity violation that was observed in the second chapter. This includes an explanation of the two-component theory of the neutrino and the derivation of a parity violating Hamiltonian. In the fourth section we finally reach our final destination, V-A theory, which provides a much improved phenomenological theory of $\beta$ decay.



\chapter{A New Force}

\paragraph*{}\begin{Large}$\mathcal{B}$\end{Large}y the end of the 1920's it was clear that there was something peculiar about $\beta$ decay. Coming off of a decade which had seen tremendous success in the development of quantum mechanics, physicists were now faced with another perplexing conundrum. Experimental evidence contradicted the prevailing ideas about $\beta$ decays and desperate attempts to salvage the theory had to be made.

$\beta$ rays were one of the three types of radiation observed in the final years of the nineteenth century, the others being $\alpha$ and $\gamma$ rays. These forms of radiation were indicative of different transformations of their mother nuclei: $\alpha$ and $\beta$ rays were emitted by nuclei which changed their electric charge (and thus transformed into the nuclei of a different element) while $\gamma$ rays were emitted by excited nuclei in order to drop to their ground state\footnote{This is analagous to the process by which electrons drop to the ground state in the system of atomic orbitals.}. The energy spectra of $\alpha$ and $\gamma$ rays were discrete, i.e. these rays could only have certain fixed energies. Against all expectations $\beta$ rays were observed to have a \emph{continuous} energy spectrum by James Chadwick in 1914. This observation would weigh on the minds of physicists for many years and would eventually lead to the Fermi Theory, the first theory of the Weak force.

\section{The Energy Spectrum Problem}

\paragraph*{}Let's try to understand why the continuous energy spectrum of $\beta$ decays was such a big problem. In the early twentieth century a $\beta$ decay was thought to proceed as $N_{i}\longrightarrow N_{f}+e^{-}$ where $N_{i}$ is the initial nucleus and $N_{f}$ is the final nucleus. Let's consider the kinematics of such a decay from a frame of reference in which $N_{i}$ is at rest. Energy conservation tells us that

\begin{multline}
 M_{i}c^{2}=M_{f}c^{2}+T_{f}+M_{e}c^{2}+T_{e} \Rightarrow T_{e}=M_{i}c^{2}-M_{f}c^{2}-M_{e}c^{2}-T_{f} \\  T_{e}\approx M_{i}c^{2}-M_{f}c^{2}-M_{e}c^{2}
\end{multline}

where we make the approximation that the recoil energy of the final nucleus, $T_{f}$ is negligible \footnote{Momentum conservation, with the nonrelativistic formula $T=\frac{p^{2}}{2M}$, gives $p_{f}=p_{e} \Rightarrow T_{f}=\frac{p_{e}^{2}}{2M_{f}}\approx0$ because of the large mass of the final nucleus.}. Thus, the kinetic energy of the electron is a fixed, discrete quantity with a value of $T_{e}\approx M_{i}c^{2}-M_{f}c^{2}-M_{e}c^{2}$. The observation of a continuous spectrum for the electrons from $\beta$ decay seemed to break the most precious rule of physics, energy conservation!

Although some physicists, including Niels Bohr, were willing to give up this most basic precept for this specific case, others were more measured in their approach. Certainly, with the amount of radical overthrows which had occured in the twenty-five years prior to 1930, it is not too surprising that some were able to concieve of having to abandon the principle of energy conservation. It would, in fact, take a bold proposal from one of the most extraordinary men in Physics to save it.

\section{Pauli: The Conscience of Physics}

\paragraph*{}Wolfgang Pauli, born in Vienna in 1900, would prove to be one of the most significant contributors to modern physics. His prodigious skills were evident very early on, having published a 237-page tome concerning Einstein's relativity at the age of twenty-one which, even now, is a standard text on the subject. At twenty-five, he would, in recognizing a fourth degree of freedom in the atomic spectra (later identified to be spin) and its implications (including what is now called the Pauli Exclusion Principle), lay the foundation for \emph{all} of modern chemistry. After being recommended by Einstein, Pauli would win the Nobel Prize for this work in 1945.

Pauli was a very harsh critic of both himself and his colleagues and would never let sloppy papers get past him without comment. It was this characteristic that led to him being dubbed ``The Conscience of Physics.'' There are several anecdotes which cast him in this role; perhaps the most well-known says of Pauli, ``... a friend showed him the paper of a young physicist which he suspected was not of great value but on which he wanted Pauli's views. Pauli remarked sadly, ``That's not right. It's not even wrong.\cite{peierls}" Pauli reserved this particularly harsh critisicm for ideas which it was not even possible to consider scientific because of their extraordinarily speculative or unclear nature. In the same vein, upon being asked for advice by a colleague whose papers were not noted for their quality, he responded ``I do not mind if you think slowly, but I do object when you publish more quickly than you think!\cite{cropper}"

The so-called ``Pauli effect'' was a curious phenomenon whose main consequence was the utter failure of  experimental apparatus located in the immediate vicinity of Pauli, even those which were in perfect working order moments before. This effect was so pronounced that it caused the experimental physicist Otto Stern to ban Pauli from his laboratory in spite of their close friendship. One particularly disquieting episode occured at the University of G\"{o}ttingen where an expensive measuring device abruptly stopped working. Pauli wasn't supposed to be anywhere near G\"{o}ttingen at the time so the director of the lab, upon his next meeting with Pauli, joked that at least he was innocent in this case. In fact, at the time of the incident Pauli had been travelling from Copenhagen to Z\"{u}rich and, it was later revealed, at the precise moment of the malfunction Pauli had stopped to await a connecting train in G\"{o}ttingen\cite{enz}.

Pauli enters into our story of $\beta$ decay around 1930, the year in which he would make one of the most brilliant and bold predictions of modern physics in order to save energy conservation. It was a prediction that would not be confirmed until 1956\cite{cowan} at which point Pauli would proclaim ``Everything comes to those who wait.''

\section{The Neutrino}

\paragraph*{}Pauli sent a brief letter concerning a possible solution to the continuous $\beta$ decay spectrum to be read to the participants of a conference held in T\"{u}bingen in December 1930. Pauli also used this letter to address the problem of ``wrong statistics'' in nuclei. Without going too far from our main discussion, we will briefly pause to understand this problem. 

Rutherford, in a 1920 lecture\cite{rutherford}, had described the nucleus as consisting of protons and electrons. To understand the consequences of this viewpoint we consider two characteristic features of a nucleus, its mass number $A$ and its charge number $Z$. Since the mass of the proton is much larger than the mass of the electron, $A$ is simply equal to the number of protons in the nucleus. Since the net charge of the nucleus was $Z$, there had to be exactly $A-Z$ electrons. For example, for a $^{14}_{7}$Ni nucleus with $A$=14 and $Z$=7, there would be 14 protons and 7 electrons. So the total number of fermions\footnote{Fermions, named after Enrico Fermi, are particles with half integer spin that obey Fermi-Dirac statistics and the Pauli exclusion principle. Bosons, on the other hand, have integer spin and obey Bose-Einstein statistics.} in the Nitrogen nucleus is $A+(A-Z)=2A-Z=21$. An system composed of an odd number of fermions is itself a fermion. However, it was clear from experiments that the Nitrogen nucleus and others like it behaved like bosons rather than fermions. This was referred to as the problem of ``wrong statistics''\cite{brandt}.

Now we quote Pauli's extraordinary letter, which tried to kill two birds (continuous $\beta$ spectrum and wrong statistics) with one stone, in its entirety\footnote{Translation by Kurt Riesselmann of Fermilab}. This letter's significance resides in its hesistant but prophetic prediction of a new particle which would later be found to actually be two particles, the \textit{neutron} and the \textit{neutrino}:

\begin{quote}
Dear Radioactive Ladies and Gentlemen,

As the bearer of these lines, to whom I graciously ask you to listen, will explain to you in more
detail, because of the "wrong" statistics of the N- and Li-6 nuclei and the continuous beta
spectrum, I have hit upon a desperate remedy to save the "exchange theorem" of statistics and
the law of conservation of energy. Namely, the possibility that in the nuclei there could exist
electrically neutral particles, which I will call neutrons, that have spin 1/2 and obey the exclusion
principle and that further differ from light quanta in that they do not travel with the velocity of
light. The mass of the neutrons should be of the same order of magnitude as the electron mass and
in any event not larger than 0.01 proton mass. - The continuous beta spectrum would then make
sense with the assumption that in beta decay, in addition to the electron, a neutron is emitted such
that the sum of the energies of neutron and electron is constant.

Now it is also a question of which forces act upon neutrons. For me, the most likely model for the
neutron seems to be, for wave-mechanical reasons (the bearer of these lines knows more), that the
neutron at rest is a magnetic dipole with a certain moment μ. The experiments seem to require
that the ionizing effect of such a neutron can not be bigger than the one of a gamma-ray, and then
μ is probably not allowed to be larger than $10^{-13}$cm.
But so far I do not dare to publish anything about this idea, and trustfully turn first to you, dear
radioactive people, with the question of how likely it is to find experimental evidence for such a
neutron if it would have the same or perhaps a 10 times larger ability to get through [material]
than a gamma-ray.

I admit that my remedy may seem almost improbable because one probably would have seen
those neutrons, if they exist, for a long time. But nothing ventured, nothing gained, and the
seriousness of the situation, due to the continuous structure of the beta spectrum, is illuminated by
a remark of my honored predecessor, Mr. Debye, who told me recently in Bruxelles: "Oh, It's
better not to think about this at all, like new taxes." Therefore one should seriously discuss every
way of rescue. Thus, dear radioactive people, scrutinize and judge. - Unfortunately, I cannot
personally appear in T\"{u}bingen since I am indispensable here in Z\"{u}rich because of a ball on the
night from December 6 to 7. With my best regards to you, and also to Mr. Back, your humble
servant

                                                     W. Pauli

\end{quote}

Thus we see Pauli's elegant proposal of a new particle intended to solve two of the major outstanding problems of contemporary physics, wrong statistics and the continuous $\beta$ decay energy spectrum, in one fell swoop. The first public presentation of this idea came in June 1931, at the American Physical Society meeting in Pasadena. Holding himself to his own high standards, Pauli cautiously decided against having this lecture printed. In October of the same year he travelled to Rome for an international congress on Nuclear Physics. It was here that Pauli's hypothesis caught the attention of Enrico Fermi, who, in the words of Pauli\cite{pauli:writings}, ``at once showed a lively interest in my idea and a very positive attitude towards my new neutral particles''.

In the next year, 1932, James Chadwick discovered a new electrically neutral particle within nuclei with a mass similar to the proton. This particle, a fermion, fixed the problem of wrong statistics which had plagued nuclear physics for several years. We now call this particle the \textit{neutron}. Our current understanding of atomic nuclei is that they are composed of $A-Z$ neutrons and $Z$ protons so that there are $A$ nuclear fermions. The $^{14}_{7}$Ni nucleus examined above is now understood to have $7$ protons and $7$ neutrons, for a total of $14$ fermions indicating bosonic behavior identical to that observed in experiments.

However, as well as this new particle did in fixing wrong statistics, it did nothing to solve the continous $\beta$ decay energy spectrum problem. It became apparent that the wrong statistics and the continuous $\beta$ spectrum, two distinct problems, would not be solved through the existence of a single new particle. Instead, a separate particle would be needed to eliminate each problem: Chadwick's neutron\footnote{Rutherford was actually the first to use the term neutron, but he had used it
to mean a nucleus made up of a proton and an electron.} for the statistics problem and a new particle, dubbed the \emph{neutrino} by Fermi, for the continuous $\beta$ spectrum\cite{pauli:writings}. Pauli's original characterization of his neutrino, as given at the October 1933 Solvay conference\footnote{This was an exceptional conference the highlights of which included the unveilings of both the positron and the neutron, a clarification of nuclear structure due to Heisenberg, Dirac's presentation of a theory of the positron, and the formal introduction of Pauli's neutrino.}, is as follows \cite{solvay}:

\begin{quote}

   As for the properties of these neutral particles, the atomic weights of
the radioactive elements in particular teach us that their mass cannot
exceed the mass of the electron by a lot. To distinguish them from the
heavy neutrons Mr. Fermi has suggested the name ``neutrino.'' It is possible that the rest mass of the neutrinos equals zero, so that they have to
propagate, like the photons, with the speed of light. In any case their
penetrating power exceeds many times that of photons of the same
energy. It seems to me admissible that the neutrinos have spin $\frac{1}{2}$ and that
they obey Fermi statistics, even though experience does not provide us
with any direct proof of this hypothesis.

\end{quote}

This light, weakly interacting, neutral particle plays a crucial role in the next part of our story, the development of Fermi's theory of $\beta$ decays, the first attempt at a formulation of what would later become known as the weak force.

\section{Early Quantum Theory}

In 1930 Fermi presented a series of lectures concerning the newly developed quantum theory in a course given at the University of Michigan titled \emph{Quantum Theory of Radiation}\cite{fermi:quant}. Fermi, who had not participated in the development of quantum mechanics himself, became thorougly aquainted with it through this course. This aquaintance would play a key role in helping him to build his theory of $\beta$ decay. It would also be helpful for us to gain some minimal knowledge of these ideas before proceeding.

Large portions of Fermi's lectures are dedicated to descriptions of the brilliant work of Paul Dirac. The importance of this work is best expressed by Fermi himself in his introductory paragraph:

\begin{quote}
 Until a few years ago it had been impossible to construct a theory of radiation which could account satisfactorily both for interference phenomena and the phenomena of emission and absorption of light by matter. The first set of phenomena was interpreted by the wave theory, and the second set by the theory of light quanta. It was not until in 1927 that Dirac succeeded in constructing a quantum theory of radiation which could explain in a unified way both types of phenomena.
\end{quote}

The importance of Dirac's theory\cite{dirac} lay in its ability to explain and predict many phenomena associated with intrinsic spin, a degree of freedom introduced only two years earlier by Pauli who had himself attempted to examine this problem with only limited success\cite{pauli:spin}. In order to grasp the significance of these results, upon which most of modern physics is built, we must go back to the most fundamental building block of wave mechanics, the stationary (time-independent), one-dimensional Schr\"{o}dinger equation:

\begin{equation}\label{TISE}
\hat{H}\Psi=\left(-\frac{\hbar^{2}}{2m} \frac{\partial^{2}}{\partial x^{2}}+V\right)\Psi=E\Psi
\end{equation}

The key point concerning Schr\"{o}dinger's formulation is that he used differential operators to express physical quantities, for example:

\begin{align}\label{ops}
x& \rightarrow x&
  p& \rightarrow \frac{\hbar}{i} \frac{\partial}{\partial x}&
E& \rightarrow -\frac{\hbar}{i} \frac{\partial}{\partial t}&
\end{align}

Where $x$ is position, $p$ is momentum, and $E$ is energy. Applying the non-relativistic expression for energy $E=\frac{p^{2}}{2m}+V$ to a wavefunction $\Psi$ with the above relations gives

\begin{equation}
 E\Psi=\left(\frac{p^{2}}{2m}+V\right)\Psi=\left(-\frac{\hbar^{2}}{2m} \frac{\partial^{2}}{\partial x^{2}}+V\right)\Psi
\end{equation}

So we see that the Schr\"{o}dinger equation is just a statement of the conservation of energy!

Let's see if this same idea will work in a relativistic formulation. In relativity the energy relation is 

\begin{equation}\label{rel}
E^{2}=(pc)^{2}+(mc^{2})^{2} 
\end{equation}

Plugging in our operators which we defined above and applying them, once again, to a wavefunction $\Psi$ we obtain another key result, the Klein-Gordon equation (sometimes called the relativistic Schr\"{o}dinger equation):

\begin{equation}
\left(-\frac{\hbar^{2}}{c^{2}}\frac{\partial^{2}}{\partial t^{2}}+\hbar^{2}\frac{\partial^{2}}{\partial x^{2}}\right)\Psi=(mc)^{2}\Psi 
\end{equation}

In the convenient units $\hbar=c=1$ this equation is just

\begin{equation}
\left(-\frac{\partial^{2}}{\partial t^{2}}+\frac{\partial^{2}}{\partial x^{2}}\right)\Psi=m^{2}\Psi 
\end{equation}

Switching to the notation of special relativity, in which $t=x^{0}$, $x=x^{1}$, abbreviating the derivatives as $\partial_{0}=\frac{\partial}{\partial x^{0}}$, and employing Einstein's summation convention in which summation is implied over identical indices (e.g. $\partial_{\mu}\partial^{\mu}=\partial_{0}\partial^{0}+\partial_{1}\partial^{1}+\partial_{2}\partial^{2}+\partial_{3}\partial^{3}$)\footnote{A value $a^{\mu}$ is not exactly the same as another value $a_{\mu}$. In special relativity they are related by the \textit{Minkowski} metric $g_{\nu \mu}$ (a diagonal four by four matrix with values $g_{00}=1$ and $g_{11}=g_{22}=g_{33}=-1$) as $g^{\nu \mu}a_{\mu}=a^{\nu}$ and $g_{\nu \mu}a^{\mu}=a_{\nu}$ where $\mu, \nu \in \{0,1,2,3\}$. This nuance is what is responsible for the sign changes you may have noticed in equation \ref{kge} in the following way: $\partial_{\mu}\partial^{\mu}=g^{\mu \nu}\partial_{\nu}\partial_{\mu}=g^{\mu \mu}\partial_{\mu}^{2}=\partial_{0}^{2}-\partial_{1}^{2}-\partial_{2}^{2}-\partial_{2}^{2}$.} the relativistic Schr\"{o}dinger equation takes on the wonderfully simple form

\begin{equation}\label{kge}
 (\partial_{\mu}\partial^{\mu}+m^{2})\Psi=0
\end{equation}

This equation, attractive as it seems, suffers from a drawback so fundamental that Schr\"{o}dinger\footnote{Schr\"{o}dinger was no fan of the new quantum theory or of its early interpretations and famously said of the theory ``I don't like it, and I'm sorry I ever had anything to do with it.''} flatly refused to publish it. It is for this reason that the equation is most often known, not as the relativistic Schr\"{o}dinger equation, but as the Klein-Gordon equation, in honor of two less hesitant physicists. The problem which stayed Schr\"{o}dinger's hand was that his equation neglected spin and was thus unable to correctly model the most valuable source of contemporary experimental data, the Hydrogen spectrum. In modern times physicists use the Klein-Gordon equation when dealing with bosons and the Dirac equation, which we will get to soon, for fermions like the electron which is responsible for Hydrogen's spectrum.

The first attempt to include spin in quantum mechanics came from Jordan and Heisenberg in 1926 \cite{jordan}. Pauli was one of the many physicists who was not satisfied with this new formulation. One particularly deep complaint was that that Heisenberg and Jordan's theory treated intrinsic spin as a traditional vector quantity, when it was clear from the Hydrogen spectrum that it was something much stranger. Pauli, after struggling with this problem for some months, would develop his own theory\cite{pauli:spin} in which spin was given by a ``classically indescribable two-valuedness'' which fit into the contemporary understanding much better. Let's take a quick look at Pauli's description and how it foreshadowed Dirac's eventual coup.

We wrote the time-independent Schr\"{o}dinger equation (eq. \ref{TISE}) as
\begin{equation}
 H\Psi=E\Psi
\end{equation}
and examined it in the context of Schr\"{o}dinger's differential operators. Now we transition to the language of Heisenberg's matrix mechanics\footnote{Quantum Mechanics was initially developed in the language of matrices by Heisenberg and only later in the language of waves by Schr\"{o}dinger. Due to the greater familiarity of physicists with waves than with matrices, Schr\"{o}dinger's formulation quickly became very popular.}, in which the above is an eigenvalue equation with $H$ as the operator matrix, $\Psi$ as the eigenvector, and energy $E$ as the eigenvalue. Equations of a similar nature hold for various other physical observables, one of which is orbital angular momentum.

Just as the Hamiltonian $H$ is the operator associated with energy $E$, the operator associated with orbital angular momentum is $\mathbf{L}$, which is actually a 3-component vector. A particle with angular momentum quantum number $l$, the $z$-projection of which we call $m$, can be described by an angular wave function $Y_{lm}$ which obeys the following eigenvalue equations

\begin{subequations}\label{grp}
\begin{align}
\mathbf{L}^{2}Y_{lm}&=l(l+1)\hbar^{2}Y_{lm}\label{first}\\
L_{z}Y_{lm}&=m\hbar Y_{lm} \label{second}
\end{align}
\end{subequations}

Pauli realized that, for intrinsic spin, \emph{only} a matrix representation is possible. He defined a new operator $\mathbf{S}$ associated with spin. This new operator had three components, each of which was a $2\times 2$ \emph{Pauli} matrix multiplied by the constant $\frac{\hbar}{2}$:

\begin{align}
S_{x}=\frac{\hbar}{2}\sigma_{x} &&
S_{y}=\frac{\hbar}{2} \sigma_{y} &&
S_{z}=\frac{\hbar}{2} \sigma_{z}
\end{align}

Where
\begin{align}
\sigma_{x}=\left( \begin{array}{cc}
0 & 1 \\
1 & 0 \end{array} \right) &&
\sigma_{y}=\left( \begin{array}{cc}
0 & -i \\
i & 0 \end{array} \right) &&
\sigma_{z}=\left( \begin{array}{cc}
1 & 0 \\
0 & -1 \end{array} \right)
\end{align}

Pauli discovered that, through the use of his new operator, eigenvalue equations exactly analagous to those which were known to hold for orbital angular momentum could be shown to also be true for intrinsic spin angular momentum. Defining $s$ as the spin quantum number and $s_{z}$ as its projection onto the z-axis these eigenvalue equations are

\begin{subequations}\label{grp2}
\begin{align}
\mathbf{S}^{2}\chi_{\pm}&=s(s+1)\hbar^{2}\chi_{\pm}\label{third}\\
S_{z}\chi_{\pm}&=s_{z}\hbar \chi_{\pm} \label{fourth}
\end{align}
\end{subequations}

The eigenvectors of this equation are important two-component objects called \textit{spinors}.

\begin{align}
\chi_{+}=
\left( \begin{array}{c}
1 \\
0 \end{array} \right)&&
\chi_{-}=
\left( \begin{array}{c}
0 \\
1 \end{array} \right)
\end{align}

Pauli's theory had two main problems. First, it was non-relativistic and, therefore, was only approximately true for low velocities. Second, and perhaps even more important, was its inability to predict a number called the ``gyromagnetic ratio''. A loop of electric current, such as that produced by an electron orbiting a hydrogen nucleus, produces a magnetic dipole moment. A seperate, intrinsic, magnetic moment is  produced by the spin of the electron\footnote{These intrinsic magneitc moments are  the source of most macroscopic magnetism}. The ratio of the produced magnetic moment to the angular momentum (the gyromagnetic ratio) for the spin is twice as large as it is for the orbital angular momentum. The fact that Pauli's theory did not predict this factor of two in the gyromagnetic ratio was a major problem.

Paul Dirac drew his main inspiration from the pursuit of elegant mathematical descriptions of physical phenomena. When asked about his philosophy of physics by a student, his response, written emphatically on a chalkboard in all capital letters, was that ``Physical laws should have mathematical beauty.'' It was this principle that led Dirac to the development of what is now known as the Dirac equation, which provided a turning point in our understanding of our universe\cite{gottfried}. 

This overarching tenet of mathematical beauty suggested to Dirac that the fundamental law of quantum mechanics should be as simple as possible. He sought to achieve this simplicity by suggesting a equation which was only first order (i.e. only involving first derivatives) in all variables as opposed to the Schr\"{o}dinger equation and the Klein-Gordon equation, which are both second order in at least one variable. He suggested an equation 

\begin{equation}\label{Dirac:eqn}
\boxed{
(i\gamma^{\mu}\partial_{\mu}-m)\Psi=0
}
\end{equation}
which he found to hold for the following four $4\times 4$ ``Dirac''(or ``Gamma'') matrices, where I is the $2\times 2$ identity matrix and $\sigma_{x},\sigma_{y},$ and $\sigma_{z}$ are the Pauli matrices.

\begin{align}
\gamma^{0}=\left( \begin{array}{cc}
I & 0 \\
0 & -I \end{array} \right) &&
\gamma^{1}=\left( \begin{array}{cc}
0 & \sigma_{x} \\
-\sigma_{x} & 0 \end{array} \right) && \\
\gamma^{2}=\left( \begin{array}{cc}
0 & \sigma_{y} \\
-\sigma_{y} & 0 \end{array} \right) &&
\gamma^{3}=\left( \begin{array}{cc}
0 & \sigma_{z} \\
-\sigma_{z} & 0 \end{array} \right)
\end{align}

We define another matrix, a product of the Dirac matrices, for later use:
\begin{equation}
\gamma^{5}=i\gamma^{0}\gamma^{1}\gamma^{2}\gamma^{3}
\end{equation}
Equation \ref{Dirac:eqn} is the Dirac equation and packs a lot of information in a very compact form. It is actually a system of four coupled equations which we can see by writing it out in a more explicit way

\begin{align}
\displaystyle\sum_{k = 1}^{4}\left(\displaystyle\sum_{\mu = 0}^{3}i(\gamma^{\mu})_{jk}\partial_{\mu}-m\delta_{jk} \right)\Psi_{k}=0 && ; && j\in\{1,2,3,4\}
\end{align}
where $\delta_{jk}$ is called the Kronecker delta and is equal to one if $j=k$ and zero otherwise. The object $\Psi$ is called a Dirac spinor and is analagous to the spinor $\chi$ in \ref{grp2} except that it has four components, not two. Two of these components were understood to provide a description of the electron in its two possible spin states. The other two components, which corresponded to \emph{negative energy} states, were a mystery for a few years before it was proposed by Dirac in mid-1931 that they represent a new type of particle, an anti-electron. Sure enough, the anti-electron (i.e. the positron) was discovered just one year later in 1932 and was a momentous success for Dirac's equation\cite{brandt}.

Among the many advantages of the Dirac equation were that it was relativisitic, it correctly predicted the structure of the Hydrogen spectrum, and it predicted the factor of two in the gyromagnetic ratio. This equation remains one of the most important in all of physics and was critical to Fermi's development of his theory of $\beta$ decays.

\section{Fermi's Theory}

In October of 1933 Fermi attended the Seventh Solvay conference\footnote{The Solvay conference, first held in 1911, is the original worldwide Physics conference. Perhaps the most prominent edition of this conference occured in 1927, amidst the whirlwind development of quantum mechanics.} in Brussels. This remarkable event, which included presentations concerning the discoveries of the positron and of the neutron as well as a talk from Pauli concerning his neutrino, left Fermi with a deep impression of many ideas that he would use in the development of a theory of $\beta$ decay.


Upon returning to Rome, Fermi proposed that an electron $e^{-}$ and a neutrino $\bar{\nu}$ \footnote{Only later was it undersood that this product must be an anti-neutrino in order to conserve lepton number} are both emitted from a neutron $n$ in the process of $\beta$ decay in a manner comparable to that in which a photon is emitted from a radioactive nucleus. This latter process is possible to describe using Dirac's  theory and what, at the time, was the fledgling theory of Quantum Electrodynamics (QED). Let's very briefly examine how the analogy between the seemingly unrelated processes of photon emission and $\beta$ decay was carried out. In QED the radiative process $p\longrightarrow p+\gamma$ where $p$ is a proton is described by the following Lagrangian density \footnote{The Langrangian $L$ is the integral of the Langrangian density $\mathcal{L}$ over all space and is equal to the difference of the kinetic and potential energies of a system. In the future I will always mean Lagrangian density when I talk about the Lagrangian.}:

\begin{equation}
 \mathcal{L}=ej_{\mu}^{(em)}A^{\mu}=e\left(\bar{u}_{p}\gamma_{\mu} u_{p}\right)A^{\mu}
\end{equation}
where $e$ is the magnitude of the electric charge of the proton, $u_{p}$ \footnote{$\bar{u}_{p}=u_{p}^{*+}\gamma_{0}$, i.e. $\bar{u}_{p}$ is obtained from $u_{p}$ by taking the complex conjugate, transforming it from a column spinor into a row spinor, and multiplying it by the Dirac matrix $\gamma_{0}$.} is the Dirac spinor of the proton, $\gamma_{\mu}$ is a Dirac matrix, and $A^{\mu}$ is the Dirac spinor of the photon. The factor $j_{\mu}^{(em)}=\left(\bar{u}_{p}\gamma_{\mu} u_{p}\right)$ is the electromagnetic current of the proton. Fermi, in order to describe $\beta$ decay, replaced $j_{\mu}^{(em)}$ with a term which dealt with the transition of the neutron into a proton $j_{\mu}^{n\rightarrow p}=\left(\bar{u}_{p}\gamma_{\mu} u_{n}\right)$ and replaced the spinor of the photon $A^{\mu}$ with a term describing the production of the electron and neutrino $j_{\nu \rightarrow e}^{\mu}=\left(\bar{u}_{e}\gamma^{\mu} u_{\nu}\right)$. He also replaced the electric charge $e$ with a new coupling constant, $G$, now known as the Fermi coupling constant $G_{F}$ \cite{morii} and, in so doing, officially discovered a new force of nature. Thus, he came up with the following Lagrangian for $\beta$ decay\footnote{Fermi's original result was actually more general than this (See Chapter 3). This Lagrangian is obtained only after assuming that the close analogy with QED holds}:

\begin{equation}\label{Fermi:Lagrangian}
 \mathcal{L}_{\beta}=G_{F}j_{\mu}^{n\rightarrow p}j_{\nu \rightarrow e}^{\mu}=G_{F}\left(\bar{u}_{p}\gamma_{\mu} u_{n}\right)\left(\bar{u}_{e}\gamma^{\mu} u_{\nu}\right)
\end{equation}

We should take notice of several important attributes of this Lagrangian. First, all of the fields in \ref{Fermi:Lagrangian} are evaluated at the same point in space and time. We call such an interaction a ``contact'' interaction. Another remarkable fact is that both Dirac currents $j_{\nu \rightarrow e}^{\mu}$ and $j_{\mu}^{n\rightarrow p}$ are ``charged''(i.e. the charge of the initial state ($\nu$, $n$) is not the same as that of the final state ($e$ , $p$)). Since electromagnetic currents are always neutral this was another tantalizing clue that $\beta$ decay was an example of something completely new\cite{renton}.

The applications of Fermi's new theory were not, however, limited to the process of the $\beta^{-}$ decay ($n\rightarrow p+e^{-}+\bar{\nu}$) which we have concentrated on so far. It is a general rule that if an interaction $a\rightarrow b+c+d$ is allowed then so are $a+\bar{b}\rightarrow c+d$, $a+\bar{c}\rightarrow b+d$ and several others. Flipping a particle from one side of the equation to another and, in the process, switching it to its anti-particle, results in an allowed process. Applying this rule to the $\beta$ decay we have been discussing we see that the following are allowed processes:

\begin{align}
 e^{-}+p \rightarrow n+\nu && \text{K capture} \\
 \bar{\nu}+p \rightarrow n+e^{+} && \text{Inverse } \beta \text{ Decay}
\end{align}
All of these ``derived'' processes were also possible to describe using Fermi theory.

Fermi sent a note to \emph{Nature} which attempted to explain his new theory only to have it refused for  ``containing abstract speculations too remote from physical reality to be of interest to the readers''\cite{rasetti}. He then sent a slightly more extended article to an Italian journal\cite{fermi:italy} and finally wrote a full account of his theory in German\cite{fermi:german} in 1934\cite{brandt}. Fermi's friend and associate Emilio Segr\'{e} gives us some insight into Fermi's attitude concerning this theory\cite{segre}: ``Fermi was fully aware of the importance of his accomplishment and said that he thought he would be remembered for this paper, his best so far.''
Fermi's theory, while representing a significant step in the right direction, had some serious problems. It would take some striking experimental and theoretical work in the next couple of decades to take our understanding of the Weak force to a new level.

\chapter{Parity Violation}

\paragraph{}\begin{Large}$\mathcal{S}$\end{Large}ome natural symmetries seem so intuitively obvious that they are never really mentioned in any serious context. Prior to the 1950's, this was the attitude that much of the physics community had towards mirror-image symmetry. Dirac, when asked to comment on the downfall of this principle, expressed the attitude of many of his contemporaries\cite{polkinghorne}: ``I never said anything about it in my book.'' The discovery of the blatant violation of this symmetry by the weak force has served as yet another warning to physicists not to put too much stock in their intuition when dealing with the subatomic world. 

In physics \emph{symmetries} and \emph{conservation laws} are very closely related. The mathematician Emmy Noether, described by Albert Einstein as the most important woman in the history of Mathematics \cite{einstein}, first discovered this connection. Noether's theorem states that if a system behaves identically with regard to some type of transformation (i.e. its \emph{Lagrangian} $\mathcal{L}$ is symmetric with respect to some transformation), then there must exist a conserved quantity within the system which is associated with this symmetry.

In order to illustrate this beautiful result, we consider a simple example. We imagine a system whose behavior is identical with respect to translations in a certain spatial direction, say along the direction parallel to the x-axis of our coordinate system. In other words, the system has exactly the same behavior at the position $\mathbf{r_{1}}=(x_{1},y_{1},z_{1})$ as it does at the position $\mathbf{r_{2}}=(x_{2},y_{1},z_{1})$ which has been translated from $\mathbf{r_{1}}$ along the x axis by a distance $\Delta x=x_{2}-x_{1}$. This is equivalent to the statement that the Lagrangian of the system, which gives a complete description of its behavior, is invariant with respect to translations along the x-axis. By Noether's theorem, we know that this symmetry implies some conservation law. It turns out that the conserved quantity associated with translations along x-axis is precisely that component of the momentum which lies parallel to this axis. Similarly, the conservation of momentum in the other two directions is simply a result of the symmetry of the laws of motion (and thus of the Lagrangian) with respect to translations in those directions \cite{Schumm}. Thus, through Noether's theorem, we find that all of our cherished conservation laws are simply consequences of the symmetries of our universe. Some of the most fundamental symmetries and their associated conservation laws are listed in Table ~\ref{sym_cons}.
\begin{table}[symmetries]
\caption{Symmetries and associated Conserved Quantities} 
\label{sym_cons}
\centering                           
\begin{tabular}{l c}              
\\ \hline \hline                             
 \emph{Symmetry} &\emph{Conserved Quantity}
\\ [0.5ex]
\hline                                      
Translation in Space & Momentum \\
Translation in Time & Energy \\
Rotation in Space & Angular Momentum \\
Reflection in Space & Parity \\
 
\hline
\end{tabular}
\end{table}

A mirror reflection can be mathematically described as an operation which transforms a vector $\mathbf{r}=(x,y,z)$ 
into 
$\mathbf{r^{'}}=-\mathbf{r}=(-x,-y,-z)$. 
Symmetry of the Lagrangian of a system with respect to such a reflection implies, through Noether's theorem, the conservation of a quantity known as parity. In order to gain an understanding of parity, let's examine the symmetry properties a wavefunction 
$\Psi(\mathbf{r})$ 
can possess under a mirror reflection. We define a new operator, 
$\hat{\mathcal{P}}$ 
which transforms 
$\Psi(\mathbf{r})$
into 
$\Psi(-\mathbf{r})$,
ie. $\hat{\mathcal{P}}\Psi(\mathbf{r})=\Psi(-\mathbf{r})$.
If the system is symmetric ($\Psi(\mathbf{r})=\Psi(-\mathbf{r})$) or antisymmetric ($\Psi(\mathbf{r})=-\Psi(-\mathbf{r})$), then it obeys a simple eigenvalue equation when operated on by $\hat{\mathcal{P}}$: $\hat{\mathcal{P}}\Psi(\mathbf{r})=p\Psi(\mathbf{r})$, where $p\in \{-1,1\}$ is the \emph{parity} of the system\cite{brandt}. Positive parity is sometimes called \emph{even} and negative, \emph{odd}. This is the quantity conserved as a result of reflection symmetry.

Parity has a couple of important properties which we should mention before we proceed:

\begin{enumerate}

\label{list 1}

\item Parity is a \emph{multiplicative quantum number}. This means that the quantity that is conserved in parity conserving interactions is \emph{not} the sum of the parities of the constituent particles
$\displaystyle\sum_{j} p_j$
but is instead the \emph{product} of these constituent parities $\displaystyle\prod_{j} p_j$.\footnote{This is actually true for any operator $\hat{\mathcal{O}}$ which, when applied twice, does nothing ($\hat{\mathcal{O}}^{2}=1$). The parity operator, which performs a mirror reflection $\mathbf{r}\longrightarrow -\mathbf{r}$, if applied twice, does nothing to the original system and thus fits this paradigm.}\cite{Hamermesh}

\item There are two types of parity: that which is \emph{intrinsic} to the particle, and a spatial part, which depends on the particle's angular momentum about a given point. For example, an electron in a hydrogen atom has an intrinsic parity of +1 and a spatial parity of $(-1)^{l}$ where $l$ is the angular momentum quantum number\footnote{This spatial element of parity is actually a result of the application of the parity operator to the spherical harmonics, which describe a wavefunctions dependence on the azimuthal and polar angles}. Generally, this spatial part is given by $(-1)^{J}$ where $J=L+S$ is the sum of the orbital and intrinsic (spin) angular momentum of the particle.

\end{enumerate} 

With this knowledge in hand, we are ready to dive into our analysis of this conservation law and its eventual violation by the Weak force.

\section{Laporte's Rule}

\paragraph*{}In 1924 German-American physicist Otto Laporte, pupil of the great Arnold Sommerfeld, realized that atomic states which underwent photon absorption or emission always ended up, after the transition, in a final state of opposite parity \cite{laporte}. We can use the properties defined above to show that this rule is simply a statement of the law of conservation of parity. The hydrogen atom originally has parity given by $P_{i}=P_{proton}P_{electron}$, the product of the parities of its constituents. The intrinsic parities of both the proton and the electron are +1.\footnote{All fermions (spin $\frac{1}{2}$ particles) have even parity by convention. The proton is made up of fermions (quarks) and the electron is a fermion.} However, the electron state also has a spatial parity of $(-1)^{l}$, where $l$ is the orbital angular momentum. Thus, the total parity of the hydrogen atom is $P_{i}=(+1)(+1)(-1)^{l}=(-1)^{l}$. Using Laporte's rule, we know that the final state of an atom which has undergone photon absorption or emission will have a parity opposite to its original state. Thus, the wavefunction of the hydrogen atom will have a parity of $P_{2}=(-1)P_{i}=(-1)^{l+1}$ after the transition. Including the intrinsic parity of the photon, which is defined as -1, we see that the total parity of the system in its final state is $P_{f}=P_{2}P_{photon}=(-1)^{l+2}=(-1)^{l}=P_{i}$. Thus we see that Laporte's rule that an atom's wavefunction must change parity after photon emission or absorption is nothing more than a statement of the conservation of parity in electromagnetic interactions.

Just a few years later in 1927, Eugene Wigner provided a proof which showed that the law of conservation of parity, as expressed in Laporte's rule, was a direct consequence of the reflection symmetry of the electromagnetic force\cite{yang}. This should come as no surprise to us after our discussion of Noether's theorem, which so clearly expressed the deep connection between symmetries and conservation laws. Wigner, whom contemporaries called ``the Silent Genius'', would go on to recieve the 1963 Nobel Prize for his work on the ``discovery and application of fundamental symmetry principles''.

When the weak force was postulated in the 1930's by Fermi to explain $\beta$ decay, most physicists took for granted that this new force of nature would, just like the electromagnetic force, obey reflection symmetry. After all, it seems intuitively obvious that the mirror-image of every natural interaction should itself be an equally probable interaction. Two decades later, a groundbreaking paper by two brilliant Chinese physicists would send this ``intuitive'' understanding of the Weak force tumbling to the ground.

\section{Lee and Yang}

\paragraph*{}The invasion of mainland China by the Japanese in the early 1940's brought together two aspiring young physicists, Tsung Dao Lee and Chen Ning Yang. Lee, who had been studying in the Kweichow province was forced by the invasion to relocate to the National Southwest University in Kunming, where he would make an aquaintance with fellow student Yang. In 1946, both Lee and Yang recieved fellowships to pursue their doctorates at the University of Chicago, where they would become fast friends.

Yang spent his time as a graduate student under Enrico Fermi, one of the most esteemed physicists in the world at the time. He had a brief stint as an experimentalist which can be summed up by a taunt that some fellow graduate students devised, ``Where there was a bang, there was Yang.''\cite{bernstein} Fermi's advice to Yang on his career was ``As a young man, work on practical problems; do not worry about things of fundamental importance''\cite{segre}. Fortunately for the global physics community, Yang respectfully decided not to follow the advice of his mentor.

Lee, on the other hand, knew that he was bound to be a theorist. He would later invoke praise from J. Robert Oppenheimer, who described him as ``one of the most brilliant theoretical physicists then known.''\cite{nobel} During his time at Columbia University, where he started work in 1953, he became known for his harsh treatment of guest speakers. One of his former graduate students tells of the humiliation of one post-doc who, after uttering his first sentence, was forced by Lee to defend it for one and a half hours\cite{derman}.

In the mid 1950's these two young physicists would completely transform contemporary thought concerning Weak interactions. The first step in this transformation was taken with the resolution of a befuddling issue that had left physicsts confounded for several years. 

\section{The $\theta$-$\tau$ Puzzle}

The assumption of parity conservation was used as a tool for the derivation of particle quantum numbers  until a serious problem came to light, forcing physicists to reconsider the validity of this principle. At the time, all strange mesons\footnote{A strange meson is a quark-antiquark bound state with either the quark or the antiquark being strange} were denoted by the letter K, and were distinguished from each other by the use of lower-case greek letters\cite{brandt}. Two positively charged strange particles in particular, the $\tau^{+}$ and the $\theta^{+}$, would spark the destruction of the idea of parity conservation in Weak interactions. The following decays had been observed for these particles:
\begin{gather}
 \tau^{+} \longrightarrow \pi^{+}+\pi^{+}+\pi^{-} \\
 \theta^{+} \longrightarrow \pi^{+}+\pi^{0}
\end{gather}

Let's find the parities of the $\tau^{+}$ and the $\theta^{+}$ from these decays. First off, we know that the pions have an intrinsic parity of -1.\footnote{This result comes from the application of parity conservation to the decay of an ``atom'' composed of deuteron ($^{2}$H nucleus) and a pion to two neutrons.} Now we have to consider the spatial contribuition to parity in these decays. The $\tau$, the $\theta$, and the pions all have intrinsic spins of 0. Thus the total initial angular momentum in both of the above decays is $J=L+S=0$, where $L=0$ is the external angular momentum and $S=0$ is the intrinsic spin\footnote{We assume that the quarks and antiquarks of the mesons do not have any relative orbital angular momentum\cite{povh}.}.

In the decay of the $\theta$, the orbital angular momentum of the two pions must be equal to zero in order to conserve the total angular momentum. In this case the spatial contribution to parity is $(-1)^{L}=(-1)^{0}=+1$. The total parity of the final state is then $P=P_{\pi^{+}}P_{\pi^{0}}P_{spatial}=(-1)(-1)(+1)=+1$. Thus, by parity conservation, $P(\theta)=+1$.

In the decay of the $\tau$, on the other hand, the situation is not so simple. The total orbital angular momentum has two components: The first is given by the angular momentum between the two $\pi^{+}$. The second, by the angular momentum of the remaining $\pi^{-}$ about the center of mass of the two $\pi^{+}$. This sum must be equal to zero in order to conserve total angular momentum. This implies that these two components of the total orbital angular momentum must have the same magnitude. In this case the spatial component of parity is given by the product of the parities given by the two components discussed above $P_{spatial}=(-1)^{L}(-1)^{L}=[(-1)^{L}]^{2}=+1$. So, the total parity of the final state of the $\tau$ decay, and thus of the $\tau$ itself, is $P(\tau)=P_{\pi^{+}}P_{\pi^{+}}P_{\pi^{-}}P_{spatial}=(-1)^{3}(+1)=-1$.\cite{morii}

So the two particles have different parities. Well, it just so happens that these two particles have masses and lifetimes which are identical within very small experimental uncertainties. Why would there be two distinct particles which have all of their properties identical except for their parities, which are opposite? This was the question that Lee and Yang first attempted to address in April 1956 at the Sixth Rochester Conference on High Energy Nuclear Physics in Rochester, New York.

\section{A Symmetry in Doubt}

\paragraph*{}The solution which was originally put forward by Lee and Yang was a rather desperate one called parity doubling: Particles with odd strangeness (both mesons and hyperons\footnote{Hyperons are bound states of three quarks (Hadrons) which contain at least one strange quark but no charm or bottom quarks}) were hypothesized to come in pairs, one member of which had even parity and the other, odd. Although this solution does seem a bit contrived, it would serve the purpose of keeping the principle of parity conservation in Weak interactions safe.

One of the participants at the 1956 Rochester Conference was experimentalist Martin Block, who was sharing a room with Richard Feynman during the course of the event. Block asked Feynman why the rule of parity conservation was seen by so many as inviolable. They talked deep into the night and, by the next day, Block had convinced Feyman to bring up the possibility of parity violation at the conference, claiming that nobody would take it seriously if he himself proposed it. We read the following from the conference proceedings \cite{brandt}\cite{ballam}:

\begin{quote}
\textit{ Feynman brought up a question of Block's: Could it be that the $\theta$ and the $\tau$ are different parity states of the same particle which has no definite parity, i.e., that parity is not conserved. That is, does nature have a way of defining right or left-handedness uniquely? Yang stated that he and Lee looked into that matter without arriving at any definite conclusions... Perhaps one could say that parity conservation, or else time inversion invariance, could be violated. Perhaps the weak interactions could all come from this same source, a violation of spacetime symmetries...}
\end{quote}

The theorists were stuck and Lee and Yang would now turn to the experimentalists for help. Lee was quite outspoken on the relationship between experimentalists and theorists, as expressed in his two ``laws'' of physicists \cite{lee2}:

\begin{enumerate}
 \item Without experimentalists, theorists tend to drift.
 \item Without theorists, experimentalists tend to falter.
\end{enumerate}

Wary of ``drifting'', Lee and Yang began an analysis of prior experimental data involving strong and weak interactions to evaluate the extent to which the law of parity conservation held. They found that there was \emph{no} evidence of parity conservation in weak interactions, and very solid evidence for it in strong interactions. Yang comments in his Nobel lecture that ``The fact that parity conservation in the weak interactions was believed for so long without experimental support was very startling. But what was more startling was the prospect that a spacetime symmetry which the physicists have learned so well may be violated. This prospect did not appeal to us.''\cite{yang}.

In the same paper where they bring to light the stunning ignorance surrounding weak parity conservation, \emph{Question of Parity Conservation in Weak Interactions} \cite{lee&yang}, Lee and Yang point out that certain quantities, called \emph{pseudoscalars}, would have a non-zero average value if there was indeed parity violation in an interaction. They also suggest several specific experiments which can be done to measure these quantities. Before we get to these experiments we should talk a little about these special quantities that would prove vital to putting the final nail in the coffin of the idea of a parity conserving weak force.

Recall that parity conservation, by Noether's theorem, is associated with reflection symmetry, which performs the transformation $\mathbf{r}\longrightarrow -\mathbf{r}$, i.e. it transforms every vector to its negative counterpart. \emph{Vector} quantities change sign under reflection. Now consider angular momentum, $\mathbf{L}=\mathbf{r}\times \mathbf{p}$. Let's see how this quantity behaves under a reflection transformation: $\mathbf{L}_{reflected}=-\mathbf{r}\times -\mathbf{p}=\mathbf{L}$. So angular momentum does not switch sign under reflection like usual vectors. It is actually an example of a \emph{pseudovector}, vectors that are unchanged under a reflection transformation. Recall that intrinsic spin is just a type of angular momentum, and is thus also a pseudovector. Let's see what happens if we take the scalar product of spin $\mathbf{s}$, a pseudovector, and the unit vector in the direction of the momentum $\hat{\mathbf{p}}$, a vector.
\begin{equation}\label{helicity}
 h=\mathbf{s}\cdot \hat{\mathbf{p}}
\end{equation}

Usually scalars don't change sign under reflection but, surprisingly, we can clearly see that, since $\hat{\mathbf{p}}$ changes sign but $\mathbf{s}$ does not, $h\longrightarrow -h$. This not a usual scalar, it is actually a \emph{pseudoscalar}. This particular pseudoscalar, the projection of the spin onto the direction of a particle's momentum is particularly important and is called the \emph{helicity}\footnote{Left handed particles are defined as having a negative helicity while right handed particles are defined as having a positive helicity.} of the particle. Lee and Yang realized that if parity conservation was violated in the weak force then reflection symmetry would also be broken, implying that the average value of some pseudoscalars would not be zero in weak interactions.

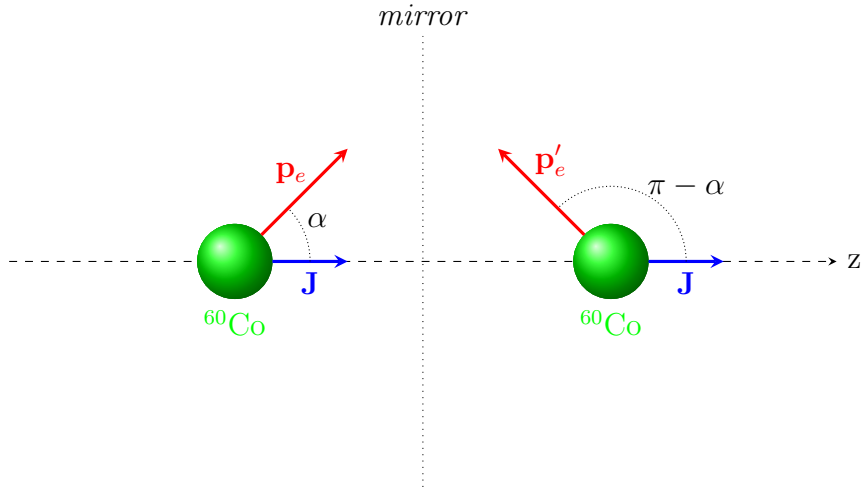
\begin{figure}
\centering
\begin{tikzpicture}[>=stealth]

\draw [->,dashed] (-3cm,0)--(8cm,0) node[anchor=west] {z};

\shade[ball color=green] (0,0) circle (.5cm);
\draw [green] (0,-.5)  node[anchor=north] {$^{60}$Co};
\draw [blue,->,very thick] (.5,0)--(1.5,0);
\draw [red,->,very thick] (.35355,.35355)--(1.5,1.5);
\draw [red] (.75,.9)  node[anchor=south] {$\mathbf{p}_{e}$};
\draw [densely dotted] (1,0) arc (0:45:1);
\draw (1.1,.55)  node {$\alpha$};
\draw [blue] (1,0)  node[anchor=north] {\textbf{J}};

\draw[dotted] (2.5cm,-3cm) -- (2.5cm,3cm) node[anchor=south] {\textit{mirror}};

\shade[ball color=green] [xshift=5cm] (0,0) circle (.5cm);
\draw [green] [xshift=5cm] (0,-.5) node[anchor=north] {$^{60}$Co};
\draw [blue,->,very thick] [xshift=5cm] (.5,0)--(1.5,0);
\draw [red,->,very thick] [xshift=5cm] (-.35355,.35355)--(-1.5,1.5);
\draw [densely dotted] [xshift=5cm] (1,0) arc (0:135:1);
\draw [xshift=5cm] (1,1)  node {$\pi-\alpha$};
\draw [red] [xshift=5cm] (-.8,1.35)  node[] {$\mathbf{p}_{e}'$};
\draw [blue] [xshift=5cm] (1,0)  node[anchor=north] {\textbf{J}};

\end{tikzpicture}
\caption{The Wu Experiment}
\label{fig:wu}
\end{figure}

The experiment suggested by Lee and Yang was based on the observation of the archetype of weak processes, $\beta$ decay. They proposed an analysis of the $^{60}$Co$(J^{p}=5^{+})\longrightarrow ^{60}$Ni$(4^{+})+e^{-}+\bar{\nu}_{e}$ decay. The pseudoscalar to be measured in this case was the projection of the electron's momentum on the spin $\mathbf{J}$ of the Cobalt nucleus. Consider a system where $\mathbf{J}$ is lined up on the $+z$ axis with a mirror at the $xy$ plane (See Figure \ref{fig:wu}). In the reflection $\mathbf{J}$ still points in the same direction as it had before since it is a pseudovector. On the other hand, the momentum of an electron $\mathbf{p}_{e}$ which had been at an angle of $\alpha$ relative to the $+z$ axis is at an angle $\pi - \alpha$ in the reflection because momentum is a vector. If this process obeys reflection symmetry (and thus parity invariance) then electrons will be just as likely to be emitted at an angle $\alpha$ as they would be at $\pi - \alpha$\cite{morii}. This is the experiment which would prove crucial to the overthrow of parity conservation.

\section{The Wu Experiment}

\paragraph*{}Even before submitting their paper to \emph{The Physical Review}, Lee and Yang had discussed their proposed $^{60}$Co decay experiment with a fellow Chinese immigrant at Columbia, Chien-Shiung Wu, known by many as Madame Wu. Madame Wu, a brilliant experimentalist with an expertise in $\beta$ decay, was so eager to conduct the experiment that she cancelled a planned trip to Europe and the Far East. 

The experiment, however, needed more than just the skill and enthusiasm of its researchers to succeed. In order to successfully align the spins of the Cobalt nuclei, their temperature had to be brought to approximately .001 Kelvin. Not many labs could achieve this feat in 1956; one that could was the Cryogenics Physics Laboratory at the National Bureau of Standards in Washington. It was to this facility that Wu turned for assistance, both in equipment and in low-temperature expertise. Ernest Ambler, who was at the National Bureau of Standards labs at the time, had developed a technique called \emph{adiabatic demagnetization} for achieving very low temperatures and was approached by Wu in June for help. Ambler had done his doctoral dissertation on the alignment of the spins of Cobalt nuclei and was very eager to offer his skills to Wu. Wu also gathered several other highly-skilled experimentalists and in October the group began to assemble and test their equipment.

On December 27, after two months of struggling with experimental apparatus, Wu's group came up with some striking results. Referring to Figure \ref{fig:wu}, it was found that electrons were much more likely to emerge at an angle $\alpha=\pi$ from the original Cobalt spin than they were to emerge along the direction of this spin ($\alpha=0$). This meant that the average value of the pseudoscalar given by the projection of the electron's momentum onto the spin of the Cobalt nucleus $\mathbf{p}_{e}\cdot \mathbf{J}$ was not equal to zero and reflection symmetry was violated \cite{wu}. Recalling the deep connections between reflection symmetry and parity conservation as described in the works of Noether and Wigner, we see that Wu's experiment attests to the violation of parity conservation. On January 9 at 2 o'clock in the morning, upon receiving confirmation of their initial results by re-running their experiment, Wu and her group brought out a bottle of Champagne and drank ``to the overthrow of parity.''\cite{forman}

The Wu experiment, although first, was not the only experimental evidence for parity violation provided at the time. Upon hearing from Lee the initial results of Wu's research, Columbia physicist Leon Lederman gathered a group to conduct a seperate experiment also suggested by Lee and Yang in their original paper. Lederman and his collaborators used Columbia's 385-MeV synchrocyclotron to analyze the $\pi-\mu-e$ decay chain and check for a similar asymmetry in the decay of the muon to that which was found by Wu in $^{60}$Co decays. This experiment provided an important independent confirmation of Wu's result \cite{lederman}.

These experiments also tell us something about the helicity of the electrons and the neutrinos involved. Recall that the spin of the Cobalt nucleus has a magnitude of $5\hbar$. The spin of the Nitrogen nucleus to which the Cobalt decays is $4\hbar$ in the same direction. The remaining $\hbar$ is carried away by the electron and the neutrino, $\frac{1}{2}\hbar$ by each. These $\frac{1}{2}\hbar$ spins of the electron and neutrino lie parallel to the spin of the original Cobalt nucleus ($\mathbf{J}$ in Figure \ref{fig:wu}) by angular momentum conservation. The fact that the electrons are emitted preferentially in a direction opposite to their spin means that their helicity ($h=\mathbf{s}\cdot \hat{\mathbf{p}}$) is more likely to be negative than positive; In other words, Wu's experiment demonstrated that electrons have a tendency to be left handed. Lee and Yang would later argue on the basis of Madame Wu's experiment that all anti-neutrinos (such as the ones emitted in the $\beta$ decay of the Cobalt nuclei) are right handed and all neutrinos are left handed.

\begin{figure}
\center
\includegraphics{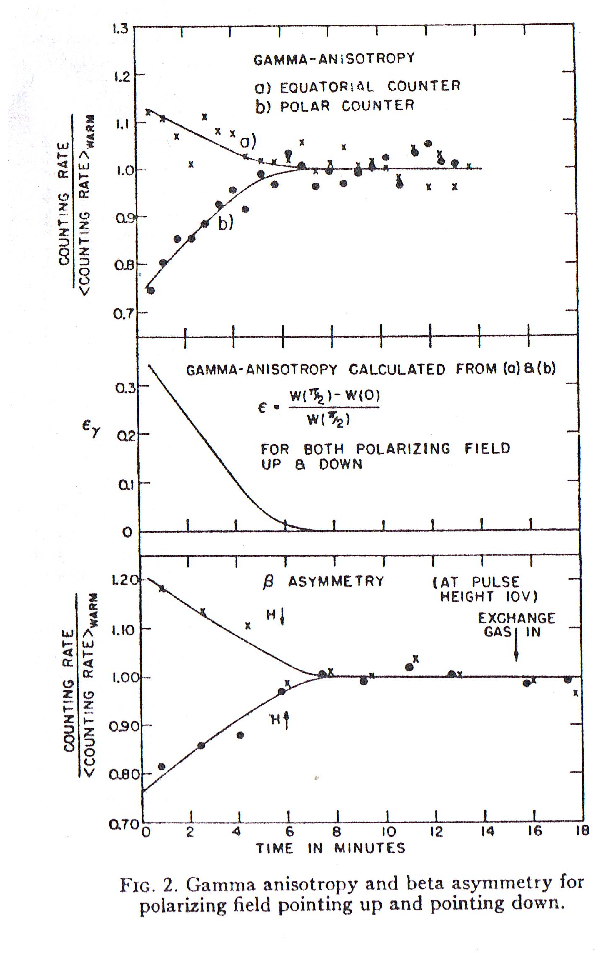}
\caption{\textit{The results of Madame Wu's experiment. Notice in particular the final plot labelled $\beta$ asymmetry. The experiment was conducted with two seperate polarization directions, thus the two curves (labelled H$\upharpoonleft$ and H$\downharpoonright$). The detector was positioned above the Cobalt source and detected many more electrons when the Cobalt was polarized down (H$\downharpoonright$) then when it was polarized up.} From \cite{wu}.}
\end{figure}

\section{A Theoretical Structure Shattered}

On January 17, prior to having heard the results of Wu's experiment, Pauli wrote a letter predicting the outcome\cite{pauli;letter}:

\begin{quote}
\textit{
I don't believe that the Lord is a weak left-hander and I am ready to bet a very high sum that the experiment will give a symmetric angular distribution of electrons. I do not see any logical connection between the strength of an interaction and its mirror invariance.
}
\end{quote}

His opinion had been shared among many great physicists of the time. Feyman, having predicted there would be no observed parity violation, lost a \$50 bet. The result was certainly a shock to the physics world. Isador Rabi said \cite{gardner}, ``A rather complete theoretical structure has been shattered at the base and we are not sure how the pieces will be put together.''

We complete our discussion of the startling discovery of parity violation with a quote from Madame Wu herself: ``One hopes that nature possesses an order that one may aspire to comprehend. When we arrive at an understanding we shall marvel how neatly all the elementary particles fit into the great scheme.'' We will see in future sections how parity violation will bring physicists closer to an understanding of this ``great scheme.''

\chapter{The Road to V-A Theory}

\paragraph{}\begin{Large}$\mathcal{I}$\end{Large}n the previous chapter we discovered, through Madame Wu's extraordinary work, that the two deeply connected principles of reflection symmetry and parity conservation are violated by the weak force. In this chapter we return to the theoretical side of our story, generalizing the framework of Fermi theory described in Chapter 1 and observing its evolution into a much more precise model in the months following the ``overthrow of parity''. Let us first turn back to Fermi's original work of the early 1930's in order to come to grips with some of its subtleties, an understanding of which will serve us well further on in this chapter.

\section{Fermi Theory Revisited}

\paragraph{}Recall that our earlier description of Fermi's theory depended heavily on the assumption that $\beta$ decay and photon emission by a nucleus are two closely related processes. Let's spend some time looking at what this assumption entails mathematically.

One of the cornerstones of modern physics is Einstein's principle of special relativity which requires that all non-gravitational\footnote{From now on we will ignore gravity since it's effects are so puny in the subatomic world we are considering.} physical laws which hold in \emph{one} inertial frame of reference\footnote{An inertial frame of reference is a non-accelerated frame. That is, one which is moving at a uniform velocity.} hold in \emph{any} inertial frame of reference. In other words, the laws of physics are symmetric under a change of inertial reference frame, which we call a Lorentz transformation. In order to express the idea of invariance under such transformations, we say that the laws of physics are \emph{Lorentz covariant}. These terms honor the great Dutch physicist, H.A. Lorentz, of whom Einstein would say: ``For me personally he meant more than all the others I have met on my life's journey.''

In Dirac's quantum field theory there are five different ways to write a physical law so that it is Lorentz covariant. These five ways are called the \emph{Dirac bilinear covariant fields} and are expressed in terms of the Dirac matrices\cite{maggiore}. Our assumption that $\beta$ decay was closely analagous to photon emission by an excited nucleus allowed us to consider only one of these bilinear forms. If we had \emph{not} made this leap in logic then we would have started with this more general expression for the Hamiltonian of $\beta$ decay as given by Fermi theory:

\begin{equation}\label{general:fermi}
 H_{w}=\displaystyle\sum_{i}\frac{G_{i}}{2}\left(\bar{u}_{p}O_{i}u_{n}\right)\left(\bar{u}_{e}O_{i}u_{\nu}\right) + \text{Hermitian Conjugate}
\end{equation}
where $O_{i}$ is one of these bilinear forms. Here are all of the possiblities, along with their names\footnote{An axial vector is just another name for a pseudovector like angular momentum which does not change sign under spatial inversion.}:

\begin{align}\label{bilin}
O_{S} & = 1 & 
	\text{Scalar (S)} & \\
O_{V} & = \gamma_{\mu} & 
	\text{Vector (V)} & \\
O_{T} & = \sigma_{\mu \nu} \equiv \frac{i}{2}\left(\gamma_{\mu}\gamma_{\nu}-\gamma_{\nu}\gamma_{\mu}\right) &
	 \text{Tensor (T)} & \\
O_{A} & = \gamma_{5}\gamma_{\mu} & 
	\text{Axial Vector (A)} & \\
O_{P} & = \gamma_{5} & 
	\text{Pseudoscalar (P)} &
\end{align}
The names of these bilinear forms come from, as you may have guessed, their properties under Lorentz transformations and reflections as discussed in the previous chapter. Our original expression, equation \ref{Fermi:Lagrangian}, set all $G_{i}=0$ for $i\in \{S, T, A, P\}$, leaving us with an expression that represented a purely vector interaction:
\begin{equation}\label{fermi:lspec}
 \mathcal{L}_{\beta}=G_{F}j_{\mu}^{n\rightarrow p}j_{\nu \rightarrow e}^{\mu}=G_{F}\left(\bar{u}_{p}\gamma_{\mu} u_{n}\right)\left(\bar{u}_{e}\gamma^{\mu} u_{\nu}\right)
\end{equation}

Fermi, having first written out the general form \ref{general:fermi}, decided to complete the analogy to QED by limiting himself to the case of a vector bilinear form \ref{fermi:lspec}. Having followed this assumption to its conclusion in Chapter 1, we will now drop it and consider the general case.

\section{Universal Fermi Interactions}

\paragraph{}So far we have only discussed a very limited set of interactions which could be described by Fermi's theory. In this section we will see how Fermi interactions became understood to be universal, leading to a clearer understanding of many processes and establishing the weak force in the pantheon of modern physics. Our story begins in the mid-1930's when a young Japanese physicist named Hideki Yukawa made a bold attempt to describe the force between nucleons through the exchange of a yet undiscovered, charged, boson\cite{yukawa}. Yukawa also made the suggestion that this new particle might be responsible for $\beta$ decay because it could decay, albeit with very small probability, into an electron-(anti)neutrino pair, the production of which was central to the Fermi theory.

Anderson, the 1932 discoverer of the positron, and his colleague, Neddermeyer, detected a particle that roughly fit Yukawa's description in 1937 \cite{anderson:muon} and, in so doing, made Yukawa, who had been relatively unknown, quite famous. The mass of the particle was just about what had been predicted by Yukawa and its decay products included an electron, just as expected. There was, however, a serious problem; Yukawa's particle should have had very strong interactions with nuclei but the detected particle was able to traverse thick layers of matter without any such interactions. In fact, this particle, which plays an important role in our story of the weak force, is the muon $\mu$ and Yukawa's meson was later found to be the pion $\pi$. The muon is the particle that will lead us from a Fermi theory which is only useful in the description of $\beta$ decay and its associated processes to a theory which encompasses many interactions, a ``universal'' Fermi theory. 

The originator of the idea of a universal Fermi theory was a young Italian physicist named Bruno Pontecorvo. Only 18 when he started working with Fermi, Pontecorvo was clearly destined for great things from an early age. While working with Irene Curie, the daughter of the great Marie Curie, in 1936 Paris, he became involved with communist groups and became exposed to political ideas that would remain with him throughout his life. During the Nazi occupation of Paris, he was unable to return to Italy due to that government's persecution of Jews like himself and instead fled to Spain. From there he went on to North America where he would spend a few years revolutionizing the oil prospecting industry and working at a nuclear research laboratory in Montreal before returning to Europe in 1949. Not surprisingly, during his time in North America he was not called upon to participate in the Manhattan project, presumably because of his affiliation to socialist groups, despite being one of the foremost global experts on nuclear physics. On August 31, 1950, during the midst of a holiday in Italy, Pontecorvo and his family abruptly fled to Stockholm, where Soviet agents helped them to defect to the USSR. It is quite natural that some media outlets have entertained speculation that he was a spy but no serious allegations in this vein have ever been voiced.

Pontecorvo noticed that the probability for the capture of muons by nuclei is similar to the analagous process involving the capture of electrons from the innermost atomic shell\footnote{This process of K capture was derived from $\beta$ decay late in Chapter 1 and could be fully described by Fermi theory} if the mass difference between the two particles is taken into account. One could describe the process of muon absorption

\begin{equation}\label{muonsab}
 \mu^{-}+p \rightarrow n+\nu
\end{equation}

using Fermi's theory. Seeing this clear connection, Pontecorvo, with the knowledge that all interactions described by the Fermi theory involved four fermions, proposed in June of 1947 that the following decays for the muon $\mu^{-}$ and its anti-particle $\mu^{+}$ were possible\cite{pontecorvo}:

\begin{align}
 \mu^{-}\rightarrow e^{-}+\nu+\bar{\nu} & & \mu^{+}\rightarrow e^{+}+\nu+\bar{\nu}
\end{align}

It is easy to check if this is really the case. Just like we saw in Chapter 1, the electron produced in such a decay should have a continuous energy spectrum. This is exactly what Jack Steinberger\footnote{Steinberger would share the 1988 Nobel Prize for the discovery of the muon neutrino, the one in eq. \ref{muonsab}}, working under Fermi in Chicago at the time, found in 1949 \cite{steinberger}. Thus, it became clear that $\beta$ decay and the other processes discussed in Chapter 1 were not the only interactions which could be described by Fermi theory. There were now a whole class of such interactions, including these muon decays, known as universal Fermi interactions.

\section{Consequences of Parity Violation}

\paragraph{}Madame Wu's 1957 experiment, which we discussed in Chapter 2, established without question that the weak force violated reflection symmetry and, thus, parity conservation. It also led us to the conclusion that electrons are preferentially left-handed (i.e. that they usually have a negative helicity). Lee and Yang, upon hearing the results of the experiment in January of 1957, postulated that all neutrinos are left-handed and all anti-neutrinos, right. This prediction was proven correct only one year later in 1958 by Goldhaber et al. \cite{goldhaber} who measured the neutrino helicity directly. Let's take a look at some of the properties of Lee and Yang's theory.

\subsection{Two Component Theory of the Neutrino}

\paragraph{}The Dirac equation, eq. \ref{Dirac:eqn}, for a neutrino is

\begin{equation}\label{nu:dirac}
 i\gamma_{\mu}\partial^{\mu}\psi-m_{\nu}\psi=0
\end{equation}

Since we are examining the helicity of the neutrino, we will work in the representation which most easily deals with helicity, the \emph{chiral}\footnote{Chirality is very nearly the same as helicity for very light particles like the neutrino.} or \emph{Weyl} basis. The Dirac matrices take on the form:

\begin{align}
\gamma^{0}=\left( \begin{array}{cc}
0 & I \\
I & 0 \end{array} \right) &&
\vec{\gamma}=\left( \begin{array}{cc}
0 & -\vec{\sigma} \\
\vec{\sigma} & 0 \end{array} \right)
\end{align}
where $\vec{\sigma}$ is Pauli's spin operator as defined in Chapter 1. The four component Dirac spinor in the chiral representation we are using takes on the form

\begin{equation}
 \psi=\left( \begin{array}{c}
\phi \\
\chi \end{array} \right)
\end{equation}
where $\phi$ and $\chi$ are two component Weyl spinors. Plugging this into \ref{nu:dirac} we have

\begin{equation}
 i\gamma_{0}\frac{\partial\psi}{\partial t}+i\vec{\gamma}\cdot \vec{\nabla}\psi-m_{\nu}\psi=0
\end{equation}
or
\begin{equation}
 i\left( \begin{array}{cc}
0 & I \\
I & 0 \end{array} \right) \left( \begin{array}{c}
\frac{\partial\phi}{\partial t} \\
\frac{\partial\chi}{\partial t} \end{array} \right)+i\left( \begin{array}{cc}
0 & -\vec{\sigma} \\
\vec{\sigma} & 0 \end{array} \right) \left( \begin{array}{c}
\vec{\nabla}\phi \\
\vec{\nabla}\chi \end{array} \right)-m_{\nu}
\left( \begin{array}{c}
\phi \\
\chi \end{array} \right)=0
\end{equation}
which, after multiplying the matrices, gives two coupled equations

\begin{align}
 i\frac{\partial\chi}{\partial t}-i\vec{\sigma}\cdot \vec{\nabla}\chi=m_{\nu}\phi \\
 i\frac{\partial\phi}{\partial t}+i\vec{\sigma}\cdot \vec{\nabla}\phi=m_{\nu}\chi
\end{align}
However, since the mass of the neutrino is either zero or negligibly small, the equations become decoupled:

\begin{align}\label{midstep}
 i\frac{\partial\chi}{\partial t}-i\vec{\sigma}\cdot \vec{\nabla}\chi=0 \\
 i\frac{\partial\phi}{\partial t}+i\vec{\sigma}\cdot \vec{\nabla}\phi=0
\end{align}
Recall from equation \ref{ops} that the energy operator and momentum operators(keep in mind that we are using the units $\hbar=c=1$) are 

\begin{align}
E \rightarrow -\frac{1}{i} \frac{\partial}{\partial t}= i\frac{\partial}{\partial t} \\
\vec{p} \rightarrow -i\vec{\nabla}
\end{align}
and from \ref{rel} that, for a massless particle like the neutrino $E\approx \left|\vec{p}\right|\equiv p$.
Using these facts, \ref{midstep} becomes
\begin{align}
 p\chi+\vec{\sigma}\cdot \vec{p}\chi=0 \\
 p\phi+-\vec{\sigma}\cdot \vec{p}\phi=0
\end{align}
or
\begin{align}\label{twocomp}
 h\chi=\frac{\vec{\sigma}\cdot \vec{p}}{p}\chi=(-1)\chi \\
 h\phi= \frac{\vec{\sigma}\cdot \vec{p}}{p}\phi=(1)\phi
\end{align}

These are eigenvalue equations for helicity (see eq. \ref{helicity}) which show that the two component spinor $\phi$ is purely right handed and that its left handed counterpart is $\chi$. The discovery that neutrinos were always left-handed was an indication that the two components of their $\phi$ spinors were equal to zero, leaving us with only two non-zero components, i.e. a \emph{two component theory} of the neutrino\cite{commins}.

\subsection{A New Lagrangian}\label{anl}

\paragraph{}Let's take a quick look back at our generalized Hamiltonian for $\beta$ decay:

\begin{equation}
 H_{w}=\displaystyle\sum_{i}\frac{G_{i}}{2}\left(\bar{u}_{p}O_{i}u_{n}\right)\left(\bar{u}_{e}O_{i}u_{\nu}\right)+\text{h.c.}
\end{equation}
Notice that the product
\begin{equation}\label{currents}
\left(\bar{u}_{p}O_{i}u_{n}\right)\left(\bar{u}_{e}O_{i}u_{\nu}\right)
\end{equation}
can take on several bilinear forms (\ref{bilin}) including the scalar product of two vectors (as in our non-generalized Lagrangian eq. \ref{fermi:lspec}), the product of two scalars, the product of two pseudosclars, or the scalar product of two axial vectors. \emph{All} of these forms are scalars, i.e. under a spatial inversion $\vec{r}\rightarrow -\vec{r}$, they remain unchanged. For example, even though a pseudoscalar like helicity changes sign under such an inversion, a product of pseudoscalars does \emph{not}; it is a scalar. Thus it becomes obvious to us that this Hamiltonian is no longer a satisfactory description of the Weak force, which we found to not be symmetric under spatial inversion.

How could we adjust the product of our currents \ref{currents} to obtain an object with odd parity? Well, we know that pseudoscalars have odd parity so let's add one (recall that the bilinear form which behaves as a pseudoscalar is $\gamma_{5}$) in order to obtain

\begin{align}
\left(\bar{u}_{p}O_{i}u_{n}\right)\left(\bar{u}_{e}O_{i}u_{\nu}\right)+\left(\bar{u}_{p}O_{i}u_{n}\right)\left(\bar{u}_{e}O_{i}C_{i}\gamma_{5}u_{\nu}\right) \\
=\left[\bar{u}_{p}O_{i}u_{n}\right]\left[\bar{u}_{e}O_{i}\left(1+C_{i}\gamma_{5}\right)u_{\nu}\right]
\end{align}
where the first term is clearly a scalar, the second term is a pseudoscalar, and $C_{i}$ is just a constant. Thus our new, parity violating, Hamiltonian is: 
\begin{equation}\label{general:Hamilton}
 H_{w}=\displaystyle\sum_{i}\frac{G_{i}}{2}\left[\bar{u}_{p}O_{i}u_{n}\right]\left[\bar{u}_{e}O_{i}\left(1+C_{i}\gamma_{5}\right)u_{\nu}\right]+\text{h.c.}
\end{equation}
All that remains for us to do is to decide which covariant terms to include (i.e. which $i$ to sum over) and to determine the constant\footnote{Actually, these ``constants'' depend on the four momentum transferred in the interaction but since this is so small in these processes our approximation is reasonable.} through experiments. 

\section{The Birth of V-A Theory}

\paragraph{}The year is 1957 and the physics community, after decades of experimental and theoretical work, is on the brink of achieving a much improved understanding of the weak force. Just as the 1956 Rochester conference had presented a venue for the revolutionary suggestion of parity violation in weak interactions, the 1957 version would bring to the fore the question of the form of the Lagrangian of the weak force. The main protagonists at this conference would be the same as they had been one year earlier: Feynman, Lee, and Yang. Lee and Yang presented, along with their two-component theory of the neutrino \cite{lee:neutrinos}, two new operators $P_{L}$ and $P_{R}$ which served to transform a full four component spinor into a two component spinor which was only left or right handed. These took the form:

\begin{align}
 P_{R}=\frac{\left(1+\gamma_{5}\right)}{2} && P_{L}=\frac{\left(1-\gamma_{5}\right)}{2}
\end{align}

Prior to their presentation, Lee and Yang had given a copy of their paper to Feynman, who, after spending one evening working through it, came back the next day with a significant extension. He used Lee and Yang's new operators to replace the full spinors of the massive leptons (electrons, muons) in weak interactions with their left handed projections, i.e. for an electron we now have

\begin{align}
 u_{e}\rightarrow \frac{\left(1-\gamma_{5}\right)}{2}u_{e} && \bar{u}_{e}\rightarrow \frac{\left(1+\gamma_{5}\right)}{2}\bar{u}_{e}
\end{align}

Plugging this into the lepton current of eq. \ref{general:Hamilton}, the Hamiltonian we demonstrated for a parity violating $\beta$ decay, and using the anticommutation relation $\{\gamma_{\mu},\gamma_{5}\}=0$\footnote{That is, $\gamma_{\mu}\gamma_{5}=-\gamma_{\mu}\gamma_{5}$} as well as the unitarity of $\gamma_{5}$\footnote{This just means that $\gamma_{5}^{2}=1$} we obtain

\begin{align}
 \bar{u}_{e}O_{i}\left(1+C_{i}\gamma_{5}\right)u_{\nu}\rightarrow \bar{u}_{e}\frac{1+\gamma_{5}}{2}O_{i}\left(1+C_{i}\gamma_{5}\right)u_{\nu} \\
=\bar{u}_{e}O_{i}\frac{1\mp\gamma_{5}}{2}\left(1+C_{i}\gamma_{5}\right)u_{\nu} \\
=\bar{u}_{e}O_{i}\left(1\mp C_{i}\right)\frac{1\mp\gamma_{5}}{2}u_{\nu}
\end{align}
where the top sign holds for $i\in\{V, A\}$ and the bottom for $i\in\{S, T, P\}$. Notice that the last term in the product represents the left-handed (top sign) or right-handed (bottom sign) projection of the neutirno spinor. But we know that neutrinos are left handed, both from experiment and from Lee and Yang's theory. Thus we must pick the top sign, which corresponds to a interaction involving only V and A covariant forms! If we pick the top sign we also know that $C_{i}=-1$ ($C_{i}\neq 1$ since this would send the whole term to zero, which would be nonsense). Thus we have the following form for our $\beta$ decay Hamiltonian \ref{general:Hamilton}:

\begin{multline}\label{VandA}
 H_{w}=\frac{G_{V}}{2}\left[\bar{u}_{p}\gamma_{\mu}u_{n}\right]\left[\bar{u}_{e}\gamma^{\mu}\left(1-\gamma_{5}\right)u_{\nu}\right] \\ + \frac{G_{A}}{2}\left[\bar{u}_{p}\gamma_{5}\gamma_{\mu}u_{n}\right]\left[\bar{u}_{e}\gamma^{\mu}\left(1-\gamma_{5}\right)u_{\nu}\right] +\text{h.c.}
\end{multline}

Interactions which proceed purely through the V term are known as Fermi transitions, perhaps in honor of his original attempts to create a purely V theory analagous to that of QED, and those that proceed through the A term are known as Gamow-Teller interactions. Gamow, a political refugee from Russia, and Teller, a Hungarian immigant who would later become a strong advocate for nuclear armament, working together in Washington D.C. at the time, had actually described such interactions in 1935, just months after the first publication of Fermi's theory. The main physical difference between the two types of interactions involves the spin of the emitted lepton pair. In Gamow-Teller interactions this spin took on a triplet state (S=1) while in Fermi interactions it was a singlet (S=0)\footnote{Imagine a vector with magnitude 1 that can only be parallel, anti-parallel, or perpendicular to the z-axis. Such a vector's possible projections onto the z-axis are clearly $1, -1, 0$. On the other hand, a vector of magnitude zero always has a projection of zero. The first case is an example of a triplet and second, a singlet.}.

The only thing that remains is to determine the constants in front of the respective terms. We won't go into the details of the numerous experiments, some extremely clever, which served this purpose. The results indicated that $G_{A}\approx -1.26G_{V}$. Defining the Fermi constant $G_{F}\equiv \frac{G_{V}}{\sqrt{2}}$ and another constant $g_{A}\equiv \left|\frac{G_{A}}{G_{V}}\right|=1.26$ we see that we can simplify eq. \ref{VandA}, the $\beta$ decay Hamiltonian, to its final form:

\begin{equation}\label{VminA}
\boxed{
H_{w}=\frac{G_{F}}{\sqrt{2}}\left[\bar{u}_{p}\gamma_{\mu}\left(1-g_{A}\gamma_{5}\right)u_{n}\right]\left[\bar{u}_{e}\gamma^{\mu}\left(1-\gamma_{5}\right)u_{\nu}\right]+\text{h.c.}
} 
\end{equation}

\section{Puzzles and Opportunities}

\paragraph{}The development of V-A theory represented significant progress in describing charged current weak interactions. However, mathematical difficulties prevent this from being a \emph{complete} theory. For that we would have to wait until the stunning accomplishments of the 1960's, which brought about an elegant unification of the Electromagnetic and Weak forces through the language of group theory. During this golden age of theoretical particle physics the breakthroughs came thick and fast and lead to one of the greatest accomplishments of modern science, the Standard Model, which gave us a delightfully simple picture of the most fundamental constituents of our world.

Now, nearly half a century after these accomplishments, theoretical particle physics is more or less stationary. Whether this is a result of speculative new ``theories of everything'' which pursue Dirac's idea of mathematical beauty without giving too much thought to emperical evidence, I will leave to the reader to decide. I only hope that we remember T.D. Lee's two ``laws'':

\begin{enumerate}
 \item Without experimentalists, theorists tend to drift.
 \item Without theorists, experimentalists tend to falter.
\end{enumerate}

Luckily, the theorists need not drift for long. A vast experiment called the Large Hadron Collider will be starting up within the next few months in Geneva and will hopefully push particle physicists a little further towards the deepest truths of our universe.

\end{document}